\documentclass[review]{elsarticle}

\usepackage{amsmath}
\usepackage{amssymb}
\usepackage{bm}

\newcommand{\beq}{\begin{equation}}
\newcommand{\eeq}{\end{equation}}

\newcommand{\ben}{\begin{eqnarray}}
\newcommand{\een}{\end{eqnarray}}
\newcommand{\benn}{\begin{eqnarray*}}
\newcommand{\eenn}{\end{eqnarray*}}

\newcommand{\nc}{\newcommand}
\nc{\bq}{\begin{equation}}
\nc{\eq}{\end{equation}}
\nc{\bqy}{\begin{eqnarray}}
\nc{\eqy}{\end{eqnarray}}
\nc{\p}{\partial}

\def\inx{\int\!\! d^3x\,}

\def\R{\mathbb{R}}
\def\J{\mathbb{J}}

\def\bfv{\mathbf{v}}

\def\bfW{\mathbf{W}}
\def\bfV{\mathbf{V}}
\def\bfU{\mathbf{U}}

\def\g{\mathfrak{g}}
\def\k{\mathfrak{k}}

\def\de{\delta}

\def\om{\omega}

\def\cala{\mathcal{A}}

\def\cald{\mathcal{D}}

\def\calm{\mathcal{M}}

\def\calp{\mathcal{P}}

\def\calx{\mathcal{X}}

\def\calz{\mathcal{Z}}

\begin{document}

\begin{frontmatter}

\title{On the Hamiltonian formulation of incompressible ideal fluids and  magnetohydrodynamics  via Dirac's theory of constraints}

\author[cpt]{C. Chandre}
\ead{chandre@cpt.univ-mrs.fr}  
\author[ut]{P. J. Morrison}
\ead{morrison@physics.utexas.edu} 
\author[cpt]{E. Tassi}
\ead{tassi@cpt.univ-mrs.fr}
\address[cpt]{Centre de Physique Th\'eorique, CNRS -- Aix-Marseille Universit\'e, Campus de Luminy, case 907, F-13288 Marseille cedex 09, France}
\address[ut]{Department of Physics and Institute for Fusion Studies, The University of Texas at Austin, Austin, TX 78712-1060, USA}

%\date{\today}
%\baselineskip 20 pt

\begin{abstract}
The Hamiltonian structures of the incompressible ideal fluid, including entropy advection,  and magnetohydrodynamics are investigated by making use of Dirac's theory of constrained Hamiltonian systems.  A Dirac bracket for these systems is constructed by assuming a primary constraint of constant density.  The resulting bracket is seen to naturally project onto solenoidal velocity fields. 
\end{abstract}
\end{frontmatter}

%%%%%%%%%%%%%%%%%%%%%%%%%%%%%%%%%%
%%%%%%%%%%%%%%%%%%%%%%%%%%%%%%%%%%

\section{Introduction}
\label{intro}

From the early work of Lagrange \cite{lagrange} it became clear that ideal fluid systems possess the canonical Hamiltonian form when one adopts a fluid element description, the so-called Lagrangian variable description.   Because the Lagrangian description is particle-like in nature, it is amenable to action functional and Hamiltonian formulations.  However, when Eulerian variables are incorporated  the canonical Hamiltonian structure for all ideal kinetic and fluid theories is altered because the transformation from Lagrangian to Eulerian variables is not canonical.  This results in a Hamiltonian theory in terms of noncanonical Poisson brackets (see, e.g.,\ \cite{morr82,sal88,morrison98,mars02,morr05,morrison06} for review). 

The present paper concerns the proper treatment of the incompressibility constraint of fluid mechanics in the context of the Eulerian Hamiltonian theory in terms of noncanonical Poisson brackets.  We do this by applying Dirac's method for incorporating constraints in Hamiltonian theories, a central element  of which is  a  Dirac bracket.  In the past, researchers have used Dirac brackets for various reasons in fluid mechanics \cite{Sal88,vann02, morr09,chandre,morr05,flierl11}, but the first works to use it  to explicitly  enforce the incompressibility constraint for Euler's equation in three dimensions  appear to be Refs.~\cite{nguy99,nguy01,nguy09}.  Here we first  extend the work of these authors by constructing  the Dirac bracket for the ideal fluid with the inclusion of entropy advection, which allows for the inclusion of any advected quantity like salt concentration in the ocean.  This generalization reveals that Dirac brackets of the kind considered in Refs.~\cite{nguy99,nguy01,nguy09}, as well as our generalization, can be written in a considerably simplified and perspicuous form in terms of the projection operator that takes a general vector field to   a solenoidal one.   With this realization we then construct the Dirac bracket for incompressible magnetohydrodynamics (MHD), thereby making clear its Hamiltonian structure.  We present these results together by starting from the full compressible ideal MHD equations, 
\begin{eqnarray}
&& \dot{\bf v}=-{\bf v}\cdot\nabla{\bf v}-\rho^{-1}\nabla \left(\rho^2 \frac{\partial U}{\partial \rho}\right) +\rho^{-1} (\nabla \times{\bf B})\times {\bf B}\,,
\label{eq:mhd-v}\\
&& \dot{\rho}=-\nabla\cdot (\rho{\bf v})\,,
\label{eq:mhd-rho}\\
&& \dot{\bf B}=\nabla\times({\bf v}\times{\bf B})\,,
\label{eq:mhd-B}\\
&& \dot{s}=-{\bf v}\cdot\nabla s\,,
\label{eq:mhd-s}
\end{eqnarray}
where  ${\bf v}({\bf x},t)$ is  the velocity field,  $\rho({\bf x},t)$ is the mass density, ${\bf B}({\bf x},t)$ is the magnetic field,   and $s({\bf x},t)$ is the entropy per unit mass.  All of these dynamical variables are functions of  ${\bf x} \in \mathcal{U}\subset \mathbb R^3$ as well as time.  We suppose boundary conditions are such that no surface terms appear in subsequent calculations which,  e.g.,  would be the case on a periodic box or all space.  
The observables of the MHD system are functionals of these fields, denoted generically by $F[\rho,{\bf v},{\bf B},s]$.
In terms of these variables, this system has the following Hamiltonian (energy): 
\begin{equation}
\label{eqn:H}
H[\rho,{\bf v},{\bf B},s]=\int d^3x\left( \frac{1}{2}\rho{v}^2 +\rho U(\rho,s)+\frac{{B}^2}{2}\right)\,,
\end{equation}
where ${v}^2=|{\bf v}|^2$ and ${B}^2=|{\bf B}|^2$.  With the MHD noncanonical Poisson bracket of Refs.~\cite{MG80, morr82}
\begin{eqnarray}
\{F,G\}&=&-\int d^3x \left( F_\rho \nabla\cdot G_{\bf v} +F_{\bf v}\cdot\nabla G_\rho -\rho^{-1}(\nabla\times {\bf v})\cdot \left( F_{\bf v}\times G_{\bf v}\right) \right. \nonumber\\
&& \; +\rho^{-1}\nabla s \cdot \left( F_s G_{\bf v}-F_{\bf v} G_s\right)+\left( \rho^{-1}F_{\bf v} \cdot [\nabla G_{\bf B}]-\rho^{-1}G_{\bf v} \cdot [\nabla F_{\bf B}]\right)\cdot {\bf B}\nonumber\\
&& \; \left. +{\bf B}\cdot \left( [\nabla \left(\rho^{-1}F_{\bf v}\right)] \cdot G_{\bf B}- [\nabla \left(\rho^{-1}G_{\bf v}\right)] \cdot F_{\bf B}\right]\right)\,, \label{eqn:PB}
\end{eqnarray}
where $F_{\bf v}$ denotes the functional derivative of $F$ with respect to ${\bf v}$, i.e.\  $F_{\bf v}=\delta F/\delta {\bf v}$, and  the same holds for $F_s$, $F_{\bf B}$ and $F_\rho$. Here the notation ${\bf a}\cdot [M] \cdot{\bf b}= {\bf b}\cdot ({\bf a}\cdot [M])$ is a scalar explicitly given by $a_i M_{ij} b_j$ (with repeated indices summed) for any vectors ${\bf a}$ and ${\bf b}$ and any matrix (or dyad)  $[M]$.  The bracket~(\ref{eqn:PB}) with Hamiltonian~(\ref{eqn:H}) gives the MHD equations~(\ref{eq:mhd-v})--(\ref{eq:mhd-s}) in the form $\dot F=\{F,H\}$. (Assuming $\nabla\cdot {\bf B}=0$.)

The paper is organized as follows:  In Sec.~\ref{sec:dirac} we review  Dirac's formalism for constrained Hamiltonian systems.  Then, in Sec.~\ref{incMHD} this theory is used  to obtain the noncanonical  Poisson-Dirac bracket for the incompressible ideal MHD equations including entropy advection.  Here we impose a primary constraint that is a constant and uniform density and the rest follows from Dirac's algorithm.  In particular, it is seen that the corresponding secondary constraint is that the velocity field be solenoidal.  We verify that  Poisson-Dirac bracket indeed produces the correct equations of motion.  This is followed in Sec.~\ref{comp} by a detailed comparison to previous attempts at incorporating incompressibility in  Hamiltonian formulations of incompressible ideal fluids.  Finally, in Sec.~\ref{conclu},   we summarize and conclude.  The paper also has several appendices that address various issues that arise in the text.

%%%%%%%%%%%%%%%%%%%%%%%%%%%%%%%%%%
%%%%%%%%%%%%%%%%%%%%%%%%%%%%%%%%%%
 
\section{Dirac brackets}
\label{sec:dirac}

As stated above, Dirac's theory is used for the derivation of the Hamiltonian structure of Hamiltonian systems subjected to constraints.  Dirac constructed his theory in terms of canonical Poisson brackets and detailed expositions of his theory can be found in Refs.~\cite{dira50,suda74,Bhans76,sund82,mars02}.  However, it is not difficult to show that his procedure also works for noncanonical Poisson brackets (cf., e.g., an Appendix of Ref.~\cite{morr09}).   In this section, we recall a few basic facts about  Dirac brackets in infinite dimensions in the context of noncanonical Poisson brackets. 

If we impose $K$ local constraints $\Phi_\alpha({\bf x})=0$ for $\alpha=1,\ldots, K$ on a Hamiltonian system with a Hamiltonian $H$ and a Poisson bracket $\{\cdot,\cdot\}$, the Dirac bracket is obtained from the matrix $C$ defined by the Poisson brackets between the constraints,
$$
C_{\alpha \beta}({\bf x},{\bf x}')=\{\Phi_\alpha({\bf x}),\Phi_\beta ({\bf x}')\}\,, 
$$
where we note that $C_{\alpha \beta}({\bf x},{\bf x}')=-C_{\beta\alpha }({\bf x}',{\bf x})$.
If $C$ has an inverse, then the Dirac bracket is defined as follows:
\begin{equation}
\label{eqn:DB}
\{F,G\}_*=\{F,G\}-\int d^3x\int d^3x'\,  \{F,\Phi_\alpha({\bf x})\}C^{-1}_{\alpha \beta}({\bf x},{\bf x}')\{\Phi_\beta({\bf x}'),G\},
\end{equation}
where  the coefficients $C^{-1}_{\alpha \beta}({\bf x},{\bf x}')$ satisfy 
$$
\int d^3x' \, C^{-1}_{\alpha \beta}({\bf x},{\bf x}')C_{\beta\gamma}({\bf x}',{\bf x}'')=\int d^3x' \, 
C_{\alpha \beta}({\bf x},{\bf x}')C^{-1}_{\beta\gamma}({\bf x}',{\bf x}'')
=\delta_{\alpha \gamma}\delta({\bf x}-{\bf x}''),
$$
which implies $C^{-1}_{\alpha \beta}({\bf x},{\bf x}')=-C^{-1}_{\beta\alpha }({\bf x}',{\bf x})$.

This procedure is effective only when the coefficients $C^{-1}_{\alpha \beta}({\bf x},{\bf x}')$ can be found. If $C$ is not invertible, then one needs, in general, secondary constraints to determine the Dirac bracket. The secondary constraint is given by the consistency equation which states that $\dot{\Phi}_1({\bf x})=0$ for the Hamiltonian $H+\int\! d^3x\,  u({\bf x})\Phi_1({\bf x})$. This translates into 
\beq
\int d^3x\,  \{\Phi_1({\bf x}),H\}\mu({\bf x})\approx 0,
\label{eq:SC}
\eeq
for all functions $\mu$ such that
\[
\int d^3x\,  \mu({\bf x})C({\bf x},{\bf x}')=0.
\]
Here the weak equality $\approx$ stands for an equality on the manifold defined by $\Phi_1({\bf x})=0$. Equation~(\ref{eq:SC}) gives the expression which has to be satisfied by the secondary constraint.

%%%%%%%%%%%%%%%%%%%%%%%%%%%%%%%%%%
%%%%%%%%%%%%%%%%%%%%%%%%%%%%%%%%%%

\section{Dirac bracket for ideal incompressible MHD}
\label{incMHD}

To construct the Hamiltonian theory of ideal incompressible MHD,  the first (primary) constraint is chosen to be a  constant and uniform density $\rho_0$, i.e.\
$$
\Phi_1({\bf x})=\rho({\bf x})-\rho_0.
$$
However,  the Dirac procedure can be performed for the case of a nonuniform background density (see \ref{nonu}). 
Given that $C_{11}({\bf x},{\bf x}')=0$, at least one secondary constraint is needed. This secondary constraint, denoted $\Phi_2({\bf x})$, is given by $\{\Phi_1({\bf x}),H\}=0$ which leads us naturally to 
$$
\Phi_2({\bf x})=\nabla \cdot {\bf v}\,.
$$
{F}rom the Poisson bracket~(\ref{eqn:PB}), we compute the elements $C_{\alpha \beta}({\bf x},{\bf x}')$ as
\begin{eqnarray*}
&& C_{11}({\bf x},{\bf x}')=0,\\
&& C_{12}({\bf x},{\bf x}')=\Delta \delta({\bf x}-{\bf x}'),\\
&& C_{21}({\bf x},{\bf x}')=-\Delta \delta({\bf x}-{\bf x}'),\\
&& C_{22}({\bf x},{\bf x}')=\nabla \cdot \left(\rho^{-1}(\nabla\times {\bf v})\times \nabla \delta({\bf x}-{\bf x}')\right).
\end{eqnarray*}
{F}rom these expressions, we obtain the coefficients $C_{\alpha \beta}^{-1}({\bf x},{\bf x}')$ as
\begin{eqnarray*}
&& C_{11}^{-1}({\bf x},{\bf x}')=\Delta^{-1} \nabla\cdot \left( \rho^{-1}(\nabla\times {\bf v})\times \nabla \Delta^{-1}\delta({\bf x}-{\bf x}')\right),\\
&& C_{12}^{-1}({\bf x},{\bf x}')=-\Delta^{-1}\delta({\bf x}-{\bf x}'),\\
&& C_{21}^{-1}({\bf x},{\bf x}')=\Delta^{-1}\delta({\bf x}-{\bf x}'),\\
&& C_{22}^{-1}({\bf x},{\bf x}')=0,
\end{eqnarray*}
where $\Delta^{-1}$ acts on a function $f$ as $\Delta^{-1}f({\bf x})=-(4\pi)^{-1}\int d^3x' f({\bf x}')/\vert {\bf x}-{\bf x}'\vert$. 
Given the following expressions
\begin{eqnarray*}
&& \{\Phi_1({\bf x}),G\}=-\nabla \cdot G_{\bf v},\\
&& \{\Phi_2({\bf x}),G\}=-\Delta G_\rho -\nabla \cdot \left( \rho^{-1}(\nabla\times {\bf v})\times G_{\bf v}\right)+\nabla \cdot \left( \rho^{-1}\nabla s G_s\right)\\
&& \qquad \qquad \qquad -\nabla\cdot \left(\rho^{-1} [\nabla G_{\bf B}]\cdot {\bf B}\right)+\nabla \cdot \left(\rho^{-1}\nabla \cdot [{\bf B} G_{\bf B}]\right),
\end{eqnarray*}
we deduce various contributions to the Dirac bracket~(\ref{eqn:DB}):

\begin{eqnarray*}
&& \iint d^3x d^3x' \{F,\Phi_1({\bf x})\}C_{11}^{-1}({\bf x},{\bf x}')\{\Phi_1({\bf x}'),G\}=-\int d^3x \nabla \cdot F_{\bf v} \Delta^{-1}\nabla \cdot(\rho^{-1}(\nabla\times {\bf v})\times \nabla \Delta^{-1} \nabla \cdot G_{\bf v})\,,\\
&& \iint d^3x d^3x' \{F,\Phi_1({\bf x})\}C_{12}^{-1}({\bf x},{\bf x}')\{\Phi_2({\bf x}'),G\}=\int d^3 x \nabla\cdot F_{\bf v} \left( G_\rho +\Delta^{-1} \nabla\cdot (\rho^{-1}(\nabla\times {\bf v})\times G_{\bf v}) \right.\\
&& \qquad \qquad \qquad \qquad \left. -\Delta^{-1} \nabla\cdot (\rho^{-1}\nabla s G_s)
+\Delta^{-1}\nabla \cdot(\rho^{-1}[\nabla G_{\bf B}]\cdot{\bf B})-\Delta^{-1}\nabla\cdot \left( \rho^{-1} \nabla\cdot [{\bf B} G_{\bf B}]\right)\right)\,,\\
&& \iint d^3x d^3x' \{F,\Phi_2({\bf x})\}C_{21}^{-1}({\bf x},{\bf x}')\{\Phi_1({\bf x}'),G\}=-\int d^3 x  \left( F_\rho +\Delta^{-1} \nabla\cdot (\rho^{-1}(\nabla\times {\bf v})\times F_{\bf v}) \right.\\
&& \qquad \qquad \qquad \qquad \left. -\Delta^{-1} \nabla\cdot (\rho^{-1}\nabla s F_s)
+\Delta^{-1}\nabla \cdot(\rho^{-1}[\nabla F_{\bf B}]\cdot{\bf B})-\Delta^{-1}\nabla\cdot \left( \rho^{-1} \nabla\cdot [{\bf B} F_{\bf B}]\right)\right)\nabla\cdot G_{\bf v}\,. 
\end{eqnarray*}
{F}rom the contributions associated with $C_{12}^{-1}$ and $C_{21}^{-1}$, we notice that the part $-\int d^3x (F_\rho \nabla\cdot G_{\bf v}+F_{\bf v}\cdot \nabla G_\rho) $ of the Poisson bracket~(\ref{eqn:PB}) vanishes. We also notice that the terms in the Dirac bracket only involve 
\bq
\bar{G}_{\bf v}:=G_{\bf v}-\nabla \Delta^{-1}\nabla\cdot G_{\bf v}=:\mathcal{P}\cdot  G_{\bf v}\,.
\label{eq:P}
\eq  
Two equivalent expressions for ${\cal P}$ acting on a vector ${\bf a}$ are ${\cal P}\cdot {\bf a}={\bf a}-\nabla\Delta^{-1}\nabla\cdot{\bf a}=-\nabla\times (\nabla \times \Delta^{-1}{\bf a})$. The linear projection operator ${\cal P}$ acting on vectors is symmetrical, in the sense that
$$
\int d^3 x\,  {\bf a}\cdot {\cal P}\cdot {\bf b}=\int d^3x\,  {\bf b}\cdot {\cal P}\cdot {\bf a},
$$
for any vector fields ${\bf a}({\bf x})$ and ${\bf b}({\bf x})$.
In addition, it satisfies the following properties: 
\begin{eqnarray*}
&& {\cal P}^2={\cal P}\,,\qquad  {\cal P}\cdot\nabla =0\,,\qquad {\cal P}\cdot\nabla \times =\nabla \times\,,\\
&& \nabla\times {\cal P}=\nabla \times\,,
\qquad \nabla \cdot {\cal P}=0\,.
\end{eqnarray*} 
As a consequence, we notice that the functional derivatives $\bar{G}_{\bf v}$ are divergence-free, i.e.\  $\nabla \cdot \bar{G}_{\bf v}=0$.
In terms of $\bar{G}_{\bf v}$ given by Eq.~(\ref{eq:P}) the Dirac bracket is written in the following compact form:
\begin{eqnarray}
&& \{F,G\}_*=\int d^3x \left( \rho^{-1}(\nabla\times {\bf v})\cdot \left( \bar{F}_{\bf v}\times \bar{G}_{\bf v}\right)-\rho^{-1}\nabla s \cdot \left( F_s \bar{G}_{\bf v}-\bar{F}_{\bf v} G_s\right)\right.\nonumber \\
&&   \left.-\left( \rho^{-1}\bar{F}_{\bf v} \cdot [\nabla G_{\bf B}] -\rho^{-1}\bar{G}_{\bf v} \cdot [\nabla F_{\bf B}]\right)\cdot{\bf B}-{\bf B}\cdot \left( [\nabla \left(\rho^{-1}\bar{F}_{\bf v}\right)] \cdot G_{\bf B}- [\nabla \left(\rho^{-1}\bar{G}_{\bf v}\right)]\cdot  F_{\bf B}\right]\right)\,. \label{eqn:PBD}
\end{eqnarray}
Upon comparison with bracket~(\ref{eqn:PB}), we see that this bracket is precisely  that of Refs.~\cite{MG80, morr82} with the  functional derivatives ${F}_{\bf v}$ and ${G}_{\bf v}$ replaced by the divergence-free functional derivatives $\bar{F}_{\bf v}$ and $\bar{G}_{\bf v}$ according to Eq.~(\ref{eq:P}). In this procedure, the terms of the bracket~(\ref{eqn:PB}) in $F_\rho$ or $G_\rho$ disappear because $\nabla \cdot \bar{G}_{\bf v}=0$. We also note that if we drop all terms but the first in Eq.~(\ref{eqn:PBD}), then with some manipulations one can show this bracket is equivalent to the one obtained in Ref.~\cite{nguy99}, albeit in a significantly simplified and  perspicuous form, and that this term corresponds to the  bracket of Ref.~\cite{zakharov97}.

Because the Poisson bracket~(\ref{eqn:PBD}) is exactly the bracket of Ref.~\cite{MG80} with the replacement of the functional derivatives by projected functional derivatives, one wonders if one can always construct Dirac brackets by this procedure. In \ref{app:hdvf} it is shown that not all projections produce good brackets, only those that define Hamiltonian vector fields (see also \ref{app:jacproj}).

Given that $\nabla \cdot \bar{F}_{\bf v}=0$ for all observables $F$, we obtain the following family of Casimir invariants of the Poisson bracket~(\ref{eqn:PBD}): 
\[
C[s]=\int d^3x\,  f(s),
\]
where $f(s)$ is any function of the entropy, i.e.\ it commutes with all the observables, $\{C[s],G\}_*=0$ for all $G$.
This family originates from the family of Casimir invariants of the original Poisson bracket~(\ref{eqn:PB}) given by $\int\! d^3x \, \rho f(s)$, and the fact that the Dirac constraints are also Casimir invariants.  This follows since Dirac brackets built on brackets with Casimir invariants retain those invariants (cf.\ Ref.~\cite{morr09}).  As a consequence, the term $\int d^3x\, \rho U(\rho,s)$ in the Hamiltonian is now a Casimir invariant, so that it can be dropped from the Hamiltonian because it will not give any contribution to the equations of motion (contrary to the compressible fluid or compressible  MHD cases). Upon setting $\rho_0=1$,  the Hamiltonian becomes
\bq
H=\frac1{2}\int \!\! d^3x \, \left({v}^2 + {B}^2 \right)\,.
\label{eq:dham}
\eq
Therefore, the Hamiltonian theory for  ideal  Eulerian incompressible MHD is given by the bracket (\ref{eqn:PBD}) with the Hamiltonian (\ref{eq:dham}). The equations of motion follow: For the entropy $s$,  this yields 
$$
\dot{s}=\{s,H\}_*=-\bar{\bf v}\cdot \nabla s,
$$
where,  as before,  we use  the `bar' shorthand for  solenoidal quantities, i.e.\ 
$\bar{\bf v}=\calp \cdot {\bf v}$, and evidently  $\nabla\cdot \bar{\bf v}=0$.   Note $s$ can be any advected quantity such as the concentration of salt. 

Similarly, the dynamical equation for ${\bf B}$ is obtained 
$$
\dot{\bf B}=\{{\bf B},H\}_*=- \bar{\bf v}\cdot \nabla {\bf B} + {\bf B}\cdot \nabla\bar{\bf v}= \nabla \times (\bar{\bf v}\times {\bf B})\,. 
$$
The equation for ${\bf v}$ is slightly more complicated, viz.
\begin{equation}
\label{eqn:NT}
\dot{\bf v} =\{{\bf v},H\}_* =-{\cal P}\cdot[(\nabla\times {\bf v})\times \bar{\bf v}]+{\cal P}\cdot[(\nabla\times {\bf B})\times {\bf B}].
\end{equation}
In particular,  the property that $\nabla\cdot{\cal P}=0$ implies that $\nabla\cdot \dot{\bf v}=0$,  which is consistent with the constraint $\Phi_2$. We notice that the first term in Eq.~(\ref{eqn:NT}) was obtained in Ref.~\cite{nguy99}.
Since $\bar{\bf v}\approx {\bf v}$ (weak equality with the constraint $\Phi_2$), the equations for ${\bf v}$ and ${\bf B}$ becomes
\begin{eqnarray*}
 && \dot{\bf v}=-{\bf v}\cdot \nabla {\bf v}-\nabla P_c +(\nabla\times {\bf B})\times {\bf B},\\
 && \dot{\bf B}=\nabla\times ({\bf v}\times {\bf B}),
\end{eqnarray*}
where the pressure-like term $P_c$ is given by
$$
P_c:=-\frac{{v}^2}{2}-\Delta^{-1}\nabla \cdot \big((\nabla\times {\bf v})\times {\bf v}\big)+\Delta^{-1}\nabla \cdot \big((\nabla\times {\bf B})\times {\bf B}\big)\, .
$$
Given the equation for the pressure, $P_c$ is not necessarily positive.  Lastly, we point out that there is no equation for the mass density $\rho$, since it has been eliminated altogether from the theory.

The  equations obtained above correspond to the traditional equations for  incompressible MHD. It should be noted that $\nabla \cdot{\bf v}=0$ is no longer a constraint on the flow since it is a conserved quantity. Actually, it is more than a conserved quantity since it is a Casimir invariant.  If one choses an initial condition satisfying $\nabla \cdot {\bf v}\neq 0$, then this quantity will remain constant under the dynamics. 

%%%%%%%%%%%%%%%%%%%%%%%%%%%%%%%%%%
%%%%%%%%%%%%%%%%%%%%%%%%%%%%%%%%%%

 \section{Comparisons between brackets for incompressible fluids}
 \label{comp}
 
We focus now on ordinary fluids and in particular we consider different formulations for describing the motion of an ideal incompressible fluid.  
 In his famous treatise  Lagrange  \cite{lagrange} provides descriptions of both incompressible and compressible ideal fluids.  Lagrange uses what is now generally called the Lagrangian variable description,  whereby the dynamics of fluid elements, points, are treated in a spatial domain, and he constructs the Lagrangian for this infinite-dimensional system.  If we let ${\bf q}$ denote the position of a fluid element labeled by ${\bf a}$, that lies in a domain $\mathcal{U}\subset \R^3$ occupied by the fluid, then ${\bf q}\colon \mathcal{U}\rightarrow \mathcal{U}$ at each time, or ${\bf q}({\bf a},t)$.  To describe incompressible fluids,  Lagrange adds the constraint $\det|\p {\bf q}/\p {\bf a}|=1$. It naturally leads to what is now referred to as the volume preserving diffeomorphism description of the incompressible fluid.  
 A formal description of this was introduced in Refs.~\cite{arnold2,arnold3} for the Euler equations of an incompressible fluid. It was based on the fact that diffeomorphisms form an infinite parameter Lie  group with a Lie algebra given by the commutator of vector fields.  If $\cald$ denotes vector fields of $\R^3$, then the commutator  (Lie bracket) 
\begin{equation}
 [\bfV,\bfW]_L= (\bfW\cdot\nabla)\bfV -(\bfV\cdot\nabla)\bfW,
 \label{PBcom}
\end{equation} 
 is again a vector field for any $\bfV,\bfW\in \cald$, and it is an elementary exercise in vector calculus to show the Jacobi identity, $[\bfU,[\bfV,\bfW]_L ]_L+ \circlearrowleft=0$ for all $\bfU,\bfV,\bfW\in \cald$, where $\circlearrowleft$ denotes the two other terms obtained by cyclic permutation of $(\bf U,\bfV,\bfW)$.  If one restricts $\cald$ to contain only divergence-free vector fields,   
 $\bar\cald :=\{\bfV\in\cald | \nabla\cdot \bfV= 0\}$, then $\bar\cald \subset\cald$ is a 
 Lie subalgebra, as seen by another elementary vector calculation that assures closure:  $\nabla\cdot [\bfV,\bfW]_L= 0$ if $\bfV,\bfW\in\bar\cald$.  From the Lie bracket~(\ref{PBcom}) one can construct the Lie-Poisson bracket (cf.,  e.g., 
 Refs.\cite{morrison98,mars02})
 \bq
\{F,G\}_L=\inx \bfv\cdot[F_{\bfv},G_{\bfv}]_L,
\label{LP}
 \eq
which indeed satisfies the Jacobi identity for all functionals of $\bfv$, it being of the Lie-Poisson form. However, combined with the Hamiltonian   $H=\inx v^2/2$, it does not yield the correct equations of motion for incompressible fluid mechanics since $\nabla \cdot {\bf v}$ is not conserved by the flow. 
 
Another bracket for incompressible fluids was proposed in Ref.~\cite{kuzn80}:
\[
\{F,G\}_0=\int\!\! d^3x\,  {\bm \omega} \cdot\left[\left(\nabla\times F_{\bm \omega}\right)\times\left(\nabla\times G_{\bm \omega}\right)\right]\,.
\]
With this bracket and the Hamiltonian $H=\inx v^2/2$ where ${\bm \omega}=\nabla\times {\bf v}$, the equation $\nabla \times H_{\bm \omega}=\bfv=H_{\bfv}$ leads to 
\[
\frac{\p {\bm \omega}}{\p t}= \nabla\times( \bfv\times {\bm \omega}),
\]
which is the correct equation of motion for the vorticity in both compressible and incompressible barotropic fluids. 
However, two issues should be noted: (i)  this bracket does not satisfy the Jacobi identity for functionals defined on arbitrary vector fields $\om$.  This is easily seen by the following counter example:
\[
F_1=\frac1{2}\inx {\bm \omega}\cdot\hat{\bf x}\, {y^2}\,,\quad
F_2=\frac1{2}\inx {\bm \omega}\cdot\hat{\bf y}\, {z^2}\,,\,,\quad
F_3=\inx {\bm \omega}\cdot \hat{\bf z}\, x\,,
\]
which yields,
\bq
\{F_1,\{F_2,F_3\}_0\}_0 + \circlearrowleft=- \inx{\bm \omega}\cdot\nabla(yz)\neq0\,,
\label{counter}
\eq
and (ii) it is not stated how the constraint $\nabla \cdot \bfv=0$ is to be applied, and indeed the procedure of this paper  also gives the correct equation of motion for compressible barotropic fluids.  

With regards to (i), if one considers vector fields that satisfy ${\bm \omega}=\nabla\times\bfv$,  then  Eq.~(\ref{counter}) gives zero (see \ref{MKjacobi}).  Thus, one might attempt to restrict the space of functionals on which this bracket is defined  in order to get a Lie algebra realization on such functionals. Yet,  since  ${\bm \omega}=\nabla\times\bfv$, it seems natural just to use $\bfv$ as a variable to enforce the constraint $\nabla\cdot{\bm \omega}=0$.  This  leads to a bracket similar to Eq.~(\ref{LP}), namely $\{F,G\}_{\bullet}=\int d^3x\, {\bf v}\cdot [F_{\bf v},G_{\bf v}]_{\bullet}$, where 
$$
[{\bf V},{\bf W}]_{\bullet}:=\nabla \times ({\bf V}\times {\bf W})=[{\bf V},{\bf W}]_L+{\bf V}(\nabla\cdot{\bf W}) -{\bf W}(\nabla\cdot{\bf V}).
$$ 
This bracket is not of Lie-Poisson type since $[\cdot,\cdot]_{\bullet}$ does not satisfy the Jacobi identity, as can be seen from the counterexample $({\bf V}_1,{\bf V}_2,{\bf V}_3)=(xy\hat{\bf x},y\hat{\bf y},\hat{\bf z})$, giving  $[[{\bf V}_1,{\bf V}_2]_{\bullet},{\bf V}_3]_{\bullet}+\circlearrowleft =y\hat{\bf z}$. Thus $\{F,G\}_{\bullet}$ has to be discarded even though it  has the interesting property $\{\nabla \cdot {\bf v}, F\}_{\bullet}=0$ for all observables $F$.

This returns us to issue (ii) above about enforcing $\nabla\cdot \bfv=0$. The failure of Eq.~(\ref{LP}) to give the correct equations of motion can be traced to the use of $\nabla\times F_{\bm \omega}=F_{\bfv}$, which cannot be true for all functionals because $0=\nabla\cdot\nabla\times F_{\bm \omega}=\nabla\cdot F_{\bfv}\neq0$.  Note, even if $F[\bfv]$ is defined on divergence-free vector fields, it does not follow that $\nabla\cdot  F_{\bfv}=0$.  This suggests introducing 
\[
\nabla\times F_{\bm \omega}=F_{\bfv} + \Upsilon\,, 
\]
where $\Upsilon$ is chosen to enforce the constraint, i.e. 
\bq
\nabla\times F_{\bm \omega}=F_{\bfv} -\nabla \Delta^{-1} \nabla \cdot F_{\bfv}=\calp\cdot F_{\bfv} \,.
\label{consDer}
\eq
Now, inserting Eq.~(\ref{consDer}) into Eq.~(\ref{LP}), we obtain the Dirac bracket of Sec.~\ref{incMHD}. So the correct Poisson bracket for incompressible fluids can be constructed as a Lie-Poisson bracket, from a projection of the  Lie bracket $[\cdot,\cdot]_L$ as follows:
$$
[{\bf V},{\bf W}]_P:=[{\cal P}\cdot {\bf V},{\cal P}\cdot{\bf W}]_L,
$$   
where we notice an important property for verifying the Jacobi identity  is
${\cal P}\cdot [{\cal P}\cdot{\bf V},{\cal P}\cdot{\bf W}]_L=[{\cal P}\cdot{\bf V},{\cal P}\cdot{\bf W}]_L$ (cf.\ \ref{app:jacproj}).
Previously the need for the projection for the incompressible fluid was observed in Refs.~\cite{mars02,zakharov97}.    However, in light of our work, when projection is handled appropriately,   this amounts to the Dirac bracket construction   of Ref.~\cite{nguy99},  which we here generalized. 

In closing this section we make a few more remarks.   In the two-dimensional formulations of 
Refs.~\cite{morrison81a,morr82,olver82} there is no issue with projection: unlike $\{\cdot, \cdot\}_0$ and $\{\cdot, \cdot\}_{\bullet}$,  the bracket given there satisfies the Jacobi identity for all functionals of the scalar vorticity.   Also, in the compressible formulation  the density is added as a dynamical variable (cf.\  the first term of Eq.~(9) of  
Ref.~\cite{MG80}) and the variations with respect to density in the Jacobi identity
compensate the failure of Jacobi for the second term alone (see footnotes 10 and 12 of Ref.~\cite{MG80}).  Lastly we point out that care must be taken when inserting projections on functional derivatives into Poisson brackets, for the resulting Poisson bracket may not satisfy the Jacobi identity (cf.\ \ref{app:hdvf} and  \ref{app:jacproj}).

%%%%%%%%%%%%%%%%%%%%%%%%%%%%%%%%%%
%%%%%%%%%%%%%%%%%%%%%%%%%%%%%%%%%%

 \section{Conclusions}
 \label{conclu}
 
Here  we have generalized  the Dirac bracket approach of Ref.~\cite{nguy99} by including entropy advection.  This produces a Hamiltonian description of an important missing piece of the dynamics of incompressible fluids, viz.\  that of density advection.   Recall,  $\nabla\cdot {\bf v}=0$ does not imply constant $\rho$, but that $\rho$ be advected.  If one chooses $\nabla\cdot\bf v$ as the primary constraint then one does not obtain a bracket for an advected density. 
Thus,  it would appear that density advection cannot be produced by the Dirac bracket construction.  However, having done the calculation with entropy advection we observe that $\rho$ drops out of the picture and we obtain a bracket that describes advection of a quantity $s$ by a solenoidal velocity field.  Thus, if one just reinterprets $s$ as $\rho$ we obtain the missing dynamics of density advection. 

Performing the Dirac construction for MHD with density advection showed us that this approach is equivalent to direct projection of the MHD bracket of  Refs.~\cite{MG80,morr82} to solenoidal vector fields.  It is now evident how to construct brackets for a variety of incompressible models.  If it is Lie-Poisson, then one can proceed as in \ref{app:jacproj} and if is not, then one can step through the Dirac bracket construction.  In fact, the Dirac construction is more general and can be used to enforce any  compatible constraints, as seen, e.g., in \ref{nonu}.  Evidently,  Dirac brackets provide a powerful tool that extends well beyond the results of this paper.

\section*{Acknowledgments}

 We acknowledge financial support from the Agence Nationale de la Recherche (ANR GYPSI). This work was also supported by the European Community under the contract of Association between EURATOM, CEA, and the French Research Federation for fusion study. The views and opinions expressed herein do not necessarily reflect those of the European Commission.  Also,  PJM  was  supported by U.S. Dept.\ of Energy Contract \# DE-FG05-80ET-53088 and he would like to acknowledge several helpful conversations  with Norm Lebovitz and Joseph Biello. The authors also acknowledge fruitful discussions with the \'Equipe de Dynamique Nonlin\'eaire of the Centre de Physique Th\'eorique of Marseille.

\appendix

%%%%%%%%%%%%%%%%%%%%%%%%%%%%%%%%%%
%%%%%%%%%%%%%%%%%%%%%%%%%%%%%%%%%%

\section{Generalization to ideal MHD with nonuniform background density}
\label{nonu}

Suppose the density is constant but nonuniform, which might,  e.g., be an imposed  stratification caused by gravity.  It is interesting to see where such an assumption leads when one follows the Dirac construction.    To this end we assume $\Phi_1({\bf x})=\rho-\rho_0({\bf x})=0$, 
where $\rho_0$ is the time-independent background density. Proceeding as in Sec.~\ref{incMHD},  because  $\{\Phi_1({\bf x}),\Phi_1({\bf x}')\}=0$ we  obtain the secondary constraint that has  the form
\bq
\Phi_2({\bf x})=\nabla\cdot (\rho_0({\bf x}) {\bf v})=0\,.
\label{eq:con2}
\eq
Although Eq.~(\ref{eq:con2}) is valid for compressible equilibria, to justify such a constraint on physical grounds would require a mechanism for maintaining the constraint or a time scale argument of some sort.   We will not pursue this here.

{F}rom the Poisson bracket~(\ref{eqn:PB}), we compute the elements $C_{\alpha \beta}({\bf x},{\bf x}')$ as
\begin{eqnarray*}
&& C_{11}({\bf x},{\bf x}')=0,\\
&& C_{12}({\bf x},{\bf x}')={\cal A} \delta({\bf x}-{\bf x}'),\\
&& C_{21}({\bf x},{\bf x}')=-{\cal A} \delta({\bf x}-{\bf x}'),\\
&& C_{22}({\bf x},{\bf x}')=\nabla \cdot \left(\rho_0^2 \rho^{-1}(\nabla\times {\bf v})\times \nabla \delta({\bf x}-{\bf x}')\right),
\end{eqnarray*}
where ${\cal A}$ is the symmetric operator ${\cal A} f=\nabla\cdot\left(\rho_0 \nabla f\right)$. 
Provided ${\cal A}$ is invertible, we obtain the coefficients $C_{\alpha \beta}^{-1}({\bf x},{\bf x}')$ as
\begin{eqnarray*}
&& C_{11}^{-1}({\bf x},{\bf x}')={\cal A}^{-1} \nabla\cdot \left( \rho_0^2 \rho^{-1}(\nabla\times {\bf v})\times \nabla {\cal A}^{-1}\delta({\bf x}-{\bf x}')\right),\\
&& C_{12}^{-1}({\bf x},{\bf x}')=-{\cal A}^{-1}\delta({\bf x}-{\bf x}'),\\
&& C_{21}^{-1}({\bf x},{\bf x}')={\cal A}^{-1}\delta({\bf x}-{\bf x}'),\\
&& C_{22}^{-1}({\bf x},{\bf x}')=0.
\end{eqnarray*}
Given the following expressions
\begin{eqnarray*}
&& \{\Phi_1({\bf x}),G\}=-\nabla \cdot G_{\bf v},\\
&& \{\Phi_2({\bf x}),G\}=-{\cal A}G_\rho -\nabla \cdot \left( \rho_0 \rho^{-1}(\nabla\times {\bf v})\times G_{\bf v}\right)+\nabla \cdot \left(\rho_0 \rho^{-1}\nabla s G_s\right)\\
&& \qquad \qquad \qquad -\nabla\cdot \left(\rho_0\rho^{-1} [\nabla G_{\bf B}]\cdot{\bf B}\right)+\nabla \cdot \left(\rho_0\rho^{-1}\nabla \cdot [{\bf B} G_{\bf B}]\right),
\end{eqnarray*}
we deduce various contributions to the Dirac bracket~(\ref{eqn:DB}):
\begin{eqnarray*}
&& \iint d^3x d^3x' \{F,\Phi_1({\bf x})\}C_{11}^{-1}({\bf x},{\bf x}')\{\Phi_1({\bf x}'),G\}=-\int d^3x \nabla \cdot F_{\bf v} {\cal A}^{-1}\nabla \cdot(\rho_0^2\rho^{-1}(\nabla\times {\bf v})\times \nabla {\cal A}^{-1} \nabla \cdot G_{\bf v}),\\
&& \iint d^3x d^3x' \{F,\Phi_1({\bf x})\}C_{12}^{-1}({\bf x},{\bf x}')\{\Phi_2({\bf x}'),G\}=\int d^3 x \nabla\cdot F_{\bf v} \left( G_\rho +{\cal A}^{-1} \nabla\cdot (\rho_0 \rho^{-1}(\nabla\times {\bf v})\times G_{\bf v}) \right.\\
&& \qquad \qquad \qquad \qquad \left. -{\cal A}^{-1} \nabla\cdot (\rho_0 \rho^{-1}\nabla s G_s)
+{\cal A}^{-1}\nabla \cdot(\rho_0\rho^{-1}[\nabla G_{\bf B}]\cdot{\bf B})-{\cal A}^{-1}\nabla\cdot \left( \rho_0\rho^{-1} \nabla\cdot [{\bf B} G_{\bf B}]\right)\right),\\
&& \iint d^3x d^3x' \{F,\Phi_2({\bf x})\}C_{21}^{-1}({\bf x},{\bf x}')\{\Phi_1({\bf x}'),G\}=-\int d^3 x  \left( F_\rho +{\cal A}^{-1} \nabla\cdot (\rho_0\rho^{-1}(\nabla\times {\bf v})\times F_{\bf v}) \right.\\
&& \qquad \qquad \qquad \qquad \left. -{\cal A}^{-1} \nabla\cdot (\rho_0\rho^{-1}\nabla s F_s)
+{\cal A}^{-1}\nabla \cdot(\rho_0 \rho^{-1}[\nabla F_{\bf B}]\cdot{\bf B})-{\cal A}^{-1}\nabla\cdot \left(\rho_0 \rho^{-1} \nabla\cdot [{\bf B} F_{\bf B}]\right)\right)\nabla\cdot G_{\bf v}.
\end{eqnarray*}
The Dirac bracket now reads
\begin{eqnarray}
&&\{F,G\}_*=\int d^3x \left( \rho^{-1}(\nabla\times {\bf v})\cdot \left( \hat{F}_{\bf v}\times \hat{G}_{\bf v}\right)-\rho^{-1}\nabla s \cdot \left( F_s \hat{G}_{\bf v}-\hat{F}_{\bf v} G_s\right)\right.\nonumber \\
&& \left.-\left( \rho^{-1}\hat{F}_{\bf v} \cdot [\nabla G_{\bf B}]-\rho^{-1}\hat{G}_{\bf v} \cdot [\nabla F_{\bf B}]\right)\cdot {\bf B}-{\bf B}\cdot \left( [\nabla \left(\rho^{-1}\hat{F}_{\bf v}\right)]\cdot  G_{\bf B}- [\nabla \left(\rho^{-1}\hat{G}_{\bf v}\right)] \cdot F_{\bf B}\right]\right), \label{eqn:PBD2}
\end{eqnarray}
where  $\hat{F}_{\bf v}=\calp_{\cala}\cdot F_{\bf v}=F_{\bf v}-\rho_0\nabla\left( {\cal A}^{-1}\nabla \cdot F_{\bf v}\right)$. 
Observe that $\nabla \cdot {\hat F}_{\bf v}=0$ for these equations,  as was the case for the incompressible MHD. We also notice that the Dirac bracket has the same form as that for incompressible MHD, with the only difference being  divergence-free functional derivatives ${\hat F}_{\bf v}$ replacing ${\bar F}_{\bf v}$.

In the same way as in Sec.~\ref{incMHD}, one  term in the Hamiltonian corresponds to a Casimir invariant. More precisely, from the property that $\nabla\cdot \hat{F}_{\bf v}=0$ for any observable $F$, it is shown that 
$$
C[s]=\int d^3x\,  \rho f(\rho,s),
$$
is a family of Casimir invariants, where $f$ is any function of $s$ and $\rho$. Therefore the Hamiltonian is 
$$
H=\frac1{2} \int\!\! d^3x \, \left(\rho_0{v}^2 + {B}^2 \right)\,, 
$$ 
and the  internal energy $U$ plays no  role in the dynamics, just as was the case for ideal incompressible MHD.

The two dynamical equations for $s$ and ${\bf B}$ are similar than the ones for incompressible MHD, and are given by
\[
\dot{s}=-{\bf v}\cdot \nabla s \qquad {\rm and} \qquad \dot{\bf B}=\nabla\times ({\bf v}\times{\bf B})\,,
\]
since $\hat{\bf v}={\bf v}-\rho_0\rho^{-1}\nabla({\cal A}^{-1}\nabla\cdot(\rho{\bf v})) 
\approx {\bf v}$ with the secondary constraint $\Phi_2$. The dynamical equation for ${\bf v}$ becomes
$$
\dot{\bf v}=-{\bf v}\cdot\nabla {\bf v} +\rho_0^{-1}(\nabla\times{\bf B})\times{\bf B}-\nabla W_c,
$$
where the Bernoulli-like term $W_c$ is given by 
$$
W_c=-\frac{{\bf v}^2}{2}-{\cal A}^{-1}\nabla\cdot (\rho_0(\nabla\times {\bf v})\times {\bf v})+{\cal A}^{-1}\nabla\cdot ((\nabla\times {\bf B})\times {\bf B}).
$$
Again we notice that $\nabla \cdot(\rho_0{\bf v})$ is conserved by the flow since it is a Casimir invariant. 

It should be noted that the second constraint $\Phi_2$ above had a  constant background density. Another choice would be to  use the constraint $\Phi_2$ with $\rho$ replacing $\rho_0$, i.e.\  use the following set of constraints:
\[
 \Phi_1({\bf x})=\rho-\rho_0({\bf x}) \qquad {\rm and}\qquad 
 \Phi_2({\bf x})=\nabla \cdot (\rho{\bf v}).
\]
Proceeding as above, the definition of the operator ${\cal A}$ naturally becomes 
${\cal A}f=\nabla\cdot(\rho \nabla f)$, and the expression for the Dirac bracket obtained is identical to Eq.~(\ref{eqn:PBD2}) with $\tilde{F}_{\bf v}:=F_{\bf v}-\rho \nabla ({\cal A}^{-1}\nabla \cdot F_{\bf v})$,
which still satisfies $\nabla \cdot \tilde{F}_{\bf v}=0$.

%%%%%%%%%%%%%%%%%%%%%%%%%%%%%%%%%%
%%%%%%%%%%%%%%%%%%%%%%%%%%%%%%%%%%

\section{Hamiltonian-Dirac Vector Fields}
\label{app:hdvf}

Let  $\mathcal Z$ denote a phase space manifold that is a symplectic or Poisson manifold, and  is thus equipped with a bracket operation $\{\,\cdot\,,\,\cdot\, \}\colon C^{\infty}(\calz)\times C^{\infty}(\calz)\rightarrow \R$.  We suppose the bracket satisfies the usual Lie enveloping  algebra properties and can thus be written in coordinates as
\[
\{f,g\}=\frac{\p f}{\p z^a}J^{ab}\frac{\p g}{\p z^b}
\]
for functions $f,g\in C^{\infty}(\calz)$, i.e.\ $f,g\colon \calz\rightarrow \R$.  Note the bracket above is a generic Poisson bracket and may have any form or degeneracy.  Only the Lie algebra properties are required.

We impose an even number of constraints $\Phi_{\alpha}\in C^{\infty}(\calz)$, $\alpha=1,\dots,2m$, and wish to project  Hamiltonian vector fields on $\calz$, elements of $\calx(\calz)$,  to Hamiltonian vector fields that  are tangent to a submanifold $\calm:=\cap_{\alpha} \Phi_{\alpha}$,  $\calx(\calm)$. 

As is well-known elements of $\calx(\calz)$  are linear operators, in particular, the element generated by $f\in C^{\infty}(\calz)$ has  the form 
\[
L_f=-\{f,\,\cdot\,\}= J^{ab}\frac{\p f}{\p z^b}\frac{\p }{\p z^a}\,, 
\]
and the commutator of two such elements satisfies $[L_f,L_g]=-L_{\{f,g\}}$.  Thus there is an isomorphism between the Lie algebra of such linear operators and Poisson brackets.  We wish to maintain this structure for Hamiltonian vector fields projected onto $\calx(\calm)$.

To project a Cartesian vector  into a surface defined by $\phi =$constant, one uses  the normal $\nabla \phi$  to   construct the following projection operator:
\bq
\mathbf{P}:= \mathbf{I} - \frac{\nabla \phi \nabla \phi}{|\nabla\phi|^2}
\label{P}
\eq
where $\mathbf{I}$ is the identity.  Evidently $\mathbf{P}\cdot \nabla\phi\equiv 0$.  Essentially this same idea occurs in infinite dimensions in the context of Hilbert spaces and is efficacious for application in  quantum mechanics.  However,  the problem at hand differs  from these cases in that we are interested in Hamiltonian vector fields (finite or infinite) and  our manifold is symplectic  with no intrinsic notion of metric.  Thus, if we are to proceed without  adding additional structure,  we  must construct a projection operator using only the functions $\Phi_{\alpha}$ and cosymplectic form, $J$.  With Eq.~(\ref{P}) as a guide we write
\[
\calp^{a}_{{\ }b}= \delta^a_{{\ }b} - K_{\alpha\beta}\,  \frac{\p \Phi_{\alpha}}{\p z^b}
\frac{\p \Phi_{\beta}}{\p z^c} J^{ac}
\]
where $K_{\alpha\beta}$ is chosen so that Hamiltonian vector fields generated by any of the $\Phi_{\alpha}$ are projected out, i.e.\  $\calp\cdot L_{\Phi_{\alpha}}\equiv 0$ for all $\alpha$.  Now it is desired to find  such a $K_{\alpha\beta}$  in terms of the $\{\Phi_{\alpha}\}$ and $J$ alone.  Fortunately, a direct calculation reveals that the desired quantity is given by $K_{\alpha\beta}= \{\Phi_{\alpha},\Phi_{\beta}\}^{-1}$.  Thus we have achieved our goal if this inverse exists.  Assuming this is the case we obtain the following Hamiltonian projection operator:
\[
\calp^{a}_{{\ }b}= \delta^a_{{\ }b} - \{\Phi_{\alpha},\Phi_{\beta}\}^{-1}\,  \frac{\p \Phi_{\alpha}}{\p z^b}
\frac{\p \Phi_{\beta}}{\p z^c} J^{ac}\,.
\]
Evidently
\bq
\Lambda^a_{\ f}:=\calp^{a}_{{\ }b} L^{b}_{\ f} = \calp^{a}_{{\ }b} J^{bd}\frac{\p f}{\p z^d} = J^{ad}\frac{\p f}{\p z^d}
-  \{\Phi_{\alpha},\Phi_{\beta}\}^{-1}\,  \frac{\p \Phi_{\alpha}}{\p z^b}
\frac{\p \Phi_{\beta}}{\p z^c} J^{ac} J^{bd}\frac{\p f}{\p z^d} \,,
\label{Lam}
\eq
and $\Lambda_{\Phi_{\alpha}}\equiv 0$ for all $\alpha$.  Also, an elementary calculation reveals the $\calp^2=\calp$, as expected for a projection operator. 

It remains to show that the set of projected vector fields of the form  $\Lambda_f=\calp\cdot L_f$ are Hamiltonian on the constraint submanifold: 
\bq
[\Lambda_f,\Lambda_g] = - \Lambda_{\{f,g\}_*}\,,
\label{com}
\eq
for some well-defined Poisson bracket $\{f,g\}_*$. As the notation suggests this turns out to be the Dirac bracket.

It is evident from Eq.~(\ref{Lam}) that $\Lambda^a_{\ f}=-\{f,\, \cdot\,\}_*$.  Because Eq.~(\ref{com}) is satisfied for a generic Poisson bracket, then it must be true for the Dirac bracket as well.   To see that it is true for a generic Poisson bracket we write
\bqy
[L_f,L_g]&=&L_fL_g-L_gL_f= J^{bc}\p_c f \p_b(J^{rs}\p_s g\p_r)
- (f\leftrightarrow g)\nonumber\\
&=& [J^{bc} \p_b J^{rs}  \p_c f \p_s g  + J^{bc} J^{rs}  \p_c f \p_b \p_s g ]
\p_r
- (f\leftrightarrow g)\nonumber\\
&=& - [J^{rb} \p_b J^{cs}  \p_c f \p_s g  + J^{sc} J^{rb} \p_b( \p_c f \p_s g )
\nonumber\\
&=& - L_{\{f,g\}}
\,, 
\nonumber
\eqy
where $\p_b:=\p/\p z^b$  and $\p_b$ operates only on the term immediately to its right unless parenthesis are included.  In obtaining the second equality, second derivative terms canceled in the usual way,  and in obtaining the third equality,  antisymmetry, the Jacobi identity, and relabeling were used. 

All of the above can be formally  extended  to infinite dimensions (see, e.g., Ref.~\cite{morr82}) by replacing partial derivatives by functional derivatives, sums by integrals,  and  matrix multiplication by operator action. 

%%%%%%%%%%%%%%%%%%%%%%%%%%%%%%%%%%
%%%%%%%%%%%%%%%%%%%%%%%%%%%%%%%%%%

\section{Projections and Poisson brackets}
\label{app:jacproj}

Consider the general Poisson bracket, 
\[
\{F,G\}=\int \!d\mu \frac{\de F}{\de \chi}\, \J\,  \frac{\de G}{\de \chi}\,,
\]
where $\J$ is a cosymplectic operator (generally dependent on $\chi(\mu)$) that ensures  this bracket satisfies the Jacobi identity.   Now suppose $\calp$ is some projection operator, and consider
\bq
\{F,G\}=\int \!d\mu\, \calp\left( \frac{\de F}{\de \chi}\right)\, \J\, \calp\left( \frac{\de G}{\de \chi}\right) 
=
\int \!d\mu\,  \frac{\de F}{\de \chi}\, \calp^{\dagger}\J\, \calp \,  \frac{\de G}{\de \chi}\,. 
\label{Pproj}
\eq
The bracket (\ref{Pproj}) does not in general satisfy the Jacobi identity.  However, if $\calp$ is independent of $\chi$ it may. 

If  the bracket is Lie-Poisson, then  projection onto subalgebras always produces brackets that satisfy the Jacobi identity.  Consider the Lie-Poisson bracket
\[
\{F,G\}=\int d\mu\,   \frac{\de F}{\de \chi} \, \J\,  \frac{\de G}{\de \chi} 
=\langle \chi,[F_{\chi}, G_{\chi}]\rangle \,.
\]
In this construction $F_{\chi}\in \g$ where $\g$ is a Lie algebra and hence $F_{\chi}$ is a vector.   Suppose $\calp:\g\rightarrow \k$, where $\k$ is a vector subspace of $\g$. Then, (i) the bracket
\bq
\{F,G\}_{P} 
=\langle \chi,[\calp F_{\chi}, \calp G_{\chi}]\rangle\,, 
\label{Pbracket}
\eq
is defined, and (ii) it satisfies the Jacobi identity for arbitrary functionals of $\chi$, provided $\calp[\calp F_{\chi}, \calp G_{\chi}]=[\calp F_{\chi}, \calp G_{\chi}]$, which is the case if $\k$ is a subalgebra of $\g$.   This follows from the general Jacobi identity theorem proven in Ref.~\cite{morr82} or more immediately from the fact that Eq.~(\ref{Pbracket}) is a Lie-Poisson bracket for $\k$. 

%%%%%%%%%%%%%%%%%%%%%%%%%%%%%%%%%%
%%%%%%%%%%%%%%%%%%%%%%%%%%%%%%%%%%

\section{Direct proof of Jacobi identity}
\label{MKjacobi}

We consider the bracket
\begin{equation}
\label{eq:ba}
\{F,G\}=\int d^3x \, {\bm \omega}\cdot [(\nabla\times F_{\bm\omega})\times( \nabla\times G_{\bm\omega})]
\end{equation}
In order to prove the Jacobi identity for this kind of bracket, one only needs to consider the explicit dependence of the bracket  on ${\bm \omega}$  when taking the functional derivative $\delta \{F,G\}/\delta {\bm\omega}$. In what follows, we let ${\bf f}:=\nabla \times F_{\bm\omega}$.    The functional derivative of $\{F,G\}$ with respect to ${\bm \omega}$ contains three terms: one that comes from the explicit dependence of the bracket on ${\bm \omega}$ and two other terms that are the second order functional derivatives of $F$ and $G$. It has been shown in Ref.~\cite{morr82} that the only important term comes from the explicit dependence on the variables, i.e.\ on ${\bm\omega}$. So from $\{F,G\}_{\bm\omega}={\bf f}\times {\bf g}$, we get
$$
\{F,\{G,H\}\}=\int d^3x\,  {\bm\omega}\cdot ({\bf f}\times \nabla \times ({\bf g}\times{\bf h})).
$$
Since $\nabla \cdot {\bf f}=0$, this becomes
\begin{equation}
\label{eq:fgh}
\{F,\{G,H\}\}=\int d^3x \, {\bm\omega}\cdot [ {\bf f}\times ({\bf h}\cdot\nabla){\bf g}-{\bf f}\times({\bf g}\cdot\nabla){\bf h}].
\end{equation}
If $\nabla \cdot {\bm\omega}=0$, there exists a vector ${\bf v}$ such that ${\bm\omega}=\nabla\times {\bf v}$. By symmetry of the operator $\nabla \times$, we obtain terms like $\nabla \times [ {\bf f}\times ({\bf h}\cdot\nabla){\bf g}-{\bf f}\times({\bf g}\cdot\nabla){\bf h}] $, which are transformed by the identity
$$
\nabla \times [{\bf f}\times ({\bf h}\cdot \nabla){\bf g}]={\bf f}\nabla \cdot [({\bf h}\cdot\nabla){\bf g}]+[({\bf h}\cdot \nabla){\bf g}\cdot\nabla] {\bf f}-({\bf f}\cdot \nabla)[({\bf h}\cdot \nabla){\bf g}].
$$
Since $\nabla\cdot {\bf f}=0$ (for all observable $F$), the divergence terms in Eq.~(\ref{eq:fgh}) vanish, i.e.\ $\nabla \cdot [({\bf h}\cdot\nabla){\bf g}]=\nabla \cdot [({\bf g}\cdot\nabla){\bf h}]$.
In addition, from the following identity
$
({\bf f}\cdot \nabla)[({\bf h}\cdot \nabla){\bf g}]=[({\bf f}\cdot \nabla){\bf h}\cdot\nabla] {\bf g}+f_ih_j\partial_i\partial_j {\bf g}
$,
we have
$$
\nabla \times [ {\bf f}\times ({\bf h}\cdot\nabla){\bf g}-{\bf f}\times({\bf g}\cdot\nabla){\bf h}] = ({\bf h},{\bf g},{\bf f})-({\bf f},{\bf h},{\bf g})-({\bf g},{\bf h},{\bf f})+({\bf f},{\bf g},{\bf h})-f_ih_j\partial_i\partial_j {\bf g}+f_ig_j\partial_i\partial_j {\bf h},
$$
where $({\bf f},{\bf g},{\bf h}):=[({\bf f}\cdot \nabla){\bf g}\cdot\nabla] {\bf h}$. 
By adding the cyclic permutations of $F$, $G$ and $H$, we obtain
$$
\nabla \times [ {\bf f}\times ({\bf h}\cdot\nabla){\bf g}-{\bf f}\times({\bf g}\cdot\nabla){\bf h}] +\circlearrowleft=0.
$$
As a consequence,  the bracket (\ref{eq:ba}) satisfies the Jacobi identity if $\nabla\cdot{\bm \omega}=0$.

{F}rom the bracket~(\ref{eq:ba}) with $\nabla \cdot {\bm \omega}=0$, we perform the following change of variables: 
${\bm\omega}=\nabla \times {\bf v}$.   This change of variables depends on a gauge since ${\bf v}+\nabla\phi$ gives the same value for ${\bm \omega}$. For instance, if we choose $\nabla \cdot {\bf v}=0$, we obtain
$$
\nabla\times F_{\bm\omega}=F_{\bf v}-\nabla \Delta^{-1} \nabla \cdot F_{\bf v}.
$$
Thus, we end up with the Poisson bracket obtained with  Dirac's procedure for constrained Hamiltonian systems.

%%%%%%%%%%%%%%%%%%%%%%%%%%%%%%%%%%%%%%%%%%
%%%%%%%%%%%%%%%%%%%%%%%%%%%%%%%%%%%%%%%%%%%%%%%%%%%%%%%%%%%%%%%%%%%%%%%%%%%%%%%%%%%%


\begin{thebibliography}{10}

\bibitem{lagrange} J.L.~Lagrange,  M\'ecanique Analytique,  English Title:  Analytical Mechanics,  translated and edited by A.~Boissonnade and V.N.~Vagliente (Kluwer Academic, Imprint Dordrecht,  Boston, Mass. 1997). 

\bibitem{morr82}  P.J.~Morrison, in {\it Mathematical Methods in Hydrodynamics
and Integrability in Related Dynamical Systems}, AIP Conference
Proceedings {\bf 88}, edited by M. Tabor and Y. Treve (AIP, New York, 1982),
p.~13.

\bibitem{sal88} R.~Salmon,  Annu. Rev. Fluid Mech. {\bf 20}, 225 (1988).

\bibitem{morrison98} P.J.~Morrison, Rev. Mod. Phys. {\bf 70}, 467 (1998).

\bibitem{mars02} J.E.~Marsden and T.R.~Ratiu, {\it Introduction to Mechanics and Symmetry} (Springer-Verlag, Berlin, 2002).

\bibitem{morr05} P.J.~Morrison,  Phys. Plasmas {\bf 12}, 058102 (2005).

\bibitem{morrison06} P.J.~Morrison, ``Hamiltonian Fluid Mechanics'', in {\em Encyclopedia of Mathematical Physics}, vol. 2, (Elsevier, Amsterdam, 2006) p. 593.

\bibitem{Sal88} R.~Salmon, J. Fluid Mech. {\bf 196}, 345 (1988).

\bibitem{vann02} J.~Vanneste and  O.~Bokhove, Physica D {\bf 164}, 152 (2002).

\bibitem{morr09} P.J.~Morrison, N.R.~Lebovitz and J.A.~Biello, Ann. Phys. {\bf 324}, 1747 (2009).

\bibitem{chandre} C.~Chandre, E.~Tassi and  P.J.~Morrison, Phys. Plasmas  {\bf 17},  042307 (2010).

\bibitem{flierl11}  G.~Flierl and P.J.~Morrison, Physica D {\bf 240}, 212  (2011).

\bibitem{nguy99} S.~Nguyen and L.A.~Turski, Physica A {\bf 272}, 48 (1999).

\bibitem{nguy01} S.~Nguyen and  L.A.~Turski, Physica A {\bf 290}, 431 (2001).

\bibitem{nguy09} S.~Nguyen and  L.A.~Turski, Physica A {\bf 388}, 91 (2009). 

\bibitem{MG80} P.J.~Morrison and J.M.~Greene, Phys. Rev. Lett. {\bf 45}, 790 (1980); {\it ibid.}\  {\bf 48}, 569 (1982).

\bibitem{dira50} P.A.M.~Dirac, Can. J. Math. {\bf 2}, 129 (1950).

\bibitem{suda74} E.C.G.~Sudarshan and N.~Mukunda, {\it Classical Dynamics: A Modern Perspective} (John Wiley \& Sons, New York, 1974).

\bibitem{Bhans76} A.~Hanson, T.~Regge and C.~Teitelboim, {\em Constrained Hamiltonian Systems} (Accademia Nazionale dei Lincei, Roma, 1976).

\bibitem{sund82} K.~Sundermeyer, {\em Constrained Dynamics} (Springer-Verlag, Berlin, 1982).

\bibitem{zakharov97} V.E.~Zakharov and E.A.~Kuznetsov, Physics-Uspekhi {\bf 40}, 1087 (1997).

\bibitem{arnold2} V.I.~Arnold, Ann. Inst. Four. {\bf 16}, 319 (1966).

\bibitem{arnold3} V.I.~Arnold,  Usp. Mat. Nauk. [Sov. Math. Usp] {\bf 24}, 225 (1969).

\bibitem{kuzn80}  E.A.~Kuznetsov and A.V.~Mikhailov,  Phys. Lett. {\bf 77A}, 37 (1980).

\bibitem{morrison81a} P.J.~Morrison, ``Hamiltonian field description of
two-dimensional vortex fluids an guiding center plasmas'', Princeton
University Plasma Physics Laboratory Report, PPPL-1783 (1981).  Available as
American Institute of Physics Document No. PAPS-PFBPE-04-771-24, AIP
Auxiliary Publication Service, 335 East 45th Street, New York, NY 10017.

\bibitem{olver82} P.J.~Olver,  J. Math. Anal. Appl. {\bf 89}, 233  (1982). 


\end{thebibliography}
\end{document}